\documentclass[sn-nature,Numbered]{sn-jnl}

\usepackage{graphicx}%
\usepackage{multirow}%
\usepackage{amsmath,amssymb,amsfonts}%
\usepackage{amsthm}%
\usepackage{mathrsfs}%
\usepackage[title]{appendix}%
\usepackage{textcomp}%
\usepackage{manyfoot}%
\usepackage{booktabs}%
\usepackage{algorithm}%
\usepackage{algorithmicx}%
\usepackage{algpseudocode}%
\usepackage{listings}%
\usepackage{subfig}
\usepackage[dvipsnames]{xcolor}
\usepackage{wrapfig}
\usepackage{hyperref}

\newcommand{\araa}{Annu. Rev. Astron. Astrophys.}   
\newcommand{\aj}{Astron. J.}   
\newcommand{\apj}{Astrophys. J.}   
\newcommand{\apjl}{Astrophys. J. Lett.}   
\newcommand{\apjs}{Astrophys. J. Suppl. Ser.}   
\newcommand{\aap}{Astron. Astrophys.}   
\newcommand{\mnras}{Mon. Not. R. Astron. Soc.}   
\newcommand{\nat}{Nature} 
\newcommand{\pasp}{Publ. Astron. Soc. Pac.}   
\newcommand{\rmxaa}{Rev. Mexicana Astron. Astrofis.}   

\theoremstyle{thmstyleone}%
\theoremstyle{thmstyletwo}%

\theoremstyle{thmstylethree}%

\newcommand{\luv}{$\mathrm{L}_{2500\AA}$}
\newcommand{\lxray}{$\mathrm{L}_{2\mathrm{keV}}$}
\newcommand{\ledd}{$\lambda_{\mathrm{Edd}}$}
\newcommand{\lbol}{L$_{\mathrm{Bol}}$}
\newcommand{\LHa}{L$_{\mathrm{H}_{\alpha}}$}
\newcommand{\mbh}{$\mathrm{M}_{\mathrm{BH}}$}
\newcommand{\halpha}{$\mathrm{H}{\alpha}$}
\newcommand{\hbeta}{$\mathrm{H}{\beta}$}
\newcommand{\aox}{$\alpha_{\mathrm{OX}}$}
\newcommand{\nii}{\hbox{[N\,{\sc ii}]}} 
\newcommand{\ciii}{\hbox{C\,{\sc iii}]}} 
\newcommand{\jwst}{{\it JWST}}
\newcommand{\oiii}{\hbox{\sc [O\,iii]}}     
\newcommand{\civ}{\hbox{\sc C\,iv}}         
\newcommand{\lya}{\hbox{\sc Ly$\alpha$}}
\newcommand{\farcs}{$.\!\!^{\prime\prime}$}
\newcommand{\arcs}{$^{\prime\prime}$}
\newcommand{\nev}{\hbox{[Ne\,{\sc v]}}} 
\newcommand{\heii}{\hbox{He\,{\sc ii}}} 
\newcommand{\micron}{$\mu m$}

\begin{document}

\title[Article Title]{The case for super-Eddington accretion in JWST broad-line AGN during the first billion years}


\author*[1]{\fnm{Erini} \sur{Lambrides}}\email{erini.lambrides@nasa.gov}

\author[2]{\fnm{Rebecca L.} \sur{Larson}}
\equalcont{These authors contributed equally to this work.}

\author[1]{\fnm{Kristen } \sur{Garofali}}
\equalcont{These authors contributed equally to this work.}

\author[1]{\fnm{Andrew } \sur{Ptak}}

\author[3,4]{\fnm{Marco} \sur{Chiaberge}}

\author[5]{\fnm{Arianna S.} \sur{Long}}

\author[1]{\fnm{Taylor A.} \sur{Hutchison}}

\author[2,4]{\fnm{Colin} \sur{Norman}}

\author[6]{\fnm{Jed}\sur{McKinney}}


\author[6]{\fnm{Hollis B.}\sur{Akins}}

\author[6]{\fnm{Danielle A.} \sur{Berg}}

\author[6]{\fnm{John} \sur{Chisholm}}

\author[1]{\fnm{Francesca} \sur{Civano}}

\author[7]{\fnm{Aidan P.} \sur{Cloonan}}

\author[6]{\fnm{Ryan} \sur{Endsley}}

\author[8]{\fnm{Andreas L.} \sur{Faisst}}

\author[9]{\fnm{Roberto} \sur{Gilli}}

\author[10,11]{\fnm{Steven} \sur{Gillman}}

\author[12]{\fnm{Michaela} \sur{Hirschmann}}

\author[13]{\fnm{Jeyhan S.} \sur{ Kartaltepe}}

\author[14]{\fnm{Dale D.} \sur{Kocevski}}

\author[6]{\fnm{Vasily} \sur{Kokorev}}

\author[15, 16]{\fnm{Fabio} \sur{Pacucci}}

\author[17]{\fnm{Chris T.} \sur{Richardson}}

\author[2]{\fnm{Massimo} \sur{Stiavelli}}

\author[1]{\fnm{Kelly E.} \sur{Whalen}}

\affil*[1]{\orgdiv{NASA Goddard Space Flight Center}, \orgname{Code 662}, \orgaddress{\city{Greenbelt}, \postcode{20771}, \state{MD}, \country{USA}}}

\affil[2]{\orgname{Space Telescope Science Institute}, \orgaddress{\street{3700 San Martin Drive}, \city{Baltimore}, \postcode{21218}, \state{MD}, \country{ USA}}}

\affil[3]{Space Telescope Science Institute for the European Space Agency (ESA), ESA Office, 3700 San Martin Dr., Baltimore, MD, USA
}

\affil[4]{\orgdiv{The William H. Miller III Department of Physics and Astronomy}, \orgname{Johns Hopkins University}, \orgaddress{\street{3400 N. Charles Street}, \city{Baltimore}, \postcode{21218}, \state{MD}, \country{USA}}}

\affil[5]{\orgdiv{Department of Astronomy}, \orgname{University of Washington, Seattle}, \orgaddress{\street{3910 15th Ave NE}, \city{Seattle},  \state{WA} \postcode{98195-0002}, \country{USA}}}

\affil[6]{\orgdiv{Department of Astronomy}, \orgname{University of Texas at Austin}, \orgaddress{2515 Speedway, Stop C1400, \city{Austin}, \postcode{78712}, \state{TX}, \country{USA}}}

\affil[7]{\orgdiv{Department of Astronomy}, \orgname{University of Massachusetts}, \orgaddress{\street{710 North Pleasant Street}, \city{Amherst},  \state{MA} \postcode{01003}, \country{USA}}}

\affil[8]{\orgdiv{IPAC}, \orgname{California Institute of Technology}, \orgaddress{\city{Pasadena}, \postcode{91125}, \state{CA}, \country{USA}}}

\affil[9]{\orgdiv{
Osservatorio di Astrofisica e Scienza dello Spazio di Bologna}, \orgname{INAF}, \orgaddress{\street{Via P. Gobetti 93/3},  \state{Bologna} \postcode{40129}, \country{Italy}}}

\affil[10]{\orgdiv{Cosmic Dawn Center (DAWN)}, \city{Copenhagen}, \orgaddress{\country{Denmark}}}

\affil[11]{\orgdiv{DTU-Space}, \orgaddress{\street{Elektrovej  Building 328}, \city{ Kgs. Lyngby}, \postcode{2800}, \country{Denmark}}}

\affil[12]{\orgdiv{Institute of Physics}, \orgname{Ecole Polytechnique Fédérale de Lausanne (EPFL), Observatoire de Sauverny}, \orgaddress{\street{1290 Versoix}, \country{Switzerland}}}

\affil[13]{\orgdiv{Laboratory for Multiwavelength Astrophysics, School of Physics and Astronomy}, \orgname{Rochester Institute of Technology}, \orgaddress{\street{84 Lomb Memorial Drive}, \city{Rochester}, \state{NY} \postcode{14623}, \country{USA}}}

\affil[14]{\orgdiv{Department of Physics and Astronomy}, \orgname{Colby College}, \orgaddress{\city{Waterville}, \postcode{04901}, \state{ME}, \country{USA}}}

\affil[15]{\orgdiv{Center for Astrophysics}, \orgname{Harvard \& Smithsonian}, \orgaddress{\city{Cambridge}, \postcode{02138}, \state{MA}, \country{USA}}}

\affil[16]{\orgdiv{Black Hole Initiative}, \orgname{Harvard University}, \orgaddress{\city{Cambridge}, \postcode{02138}, \state{MA}, \country{USA}}}

\affil[17]{\orgdiv{Department of Physics and Astronomy}, \orgname{Elon University}, \orgaddress{\street{100 Campus Drive}, \city{Elon},  \state{NC} \postcode{27244}, \country{USA}}}

\abstract{A multitude of \jwst\ studies reveal a surprising over-abundance of over-massive accreting super-massive black holes (SMBHs) -- leading to a deepening tension between theory and observation in the first billion years of cosmic time. Across X-ray to infrared wavelengths, models built off of pre-JWST predictions fail to easily reproduce observed AGN signatures (or lack thereof), driving uncertainty around the true nature of these sources.  
Using a sample of \jwst\ AGN identified via their broadened H$\alpha$ emission and covered by the deepest X-ray surveys, we find neither any measurable X-ray emission nor any detection of high-ionization emission lines frequently associated with accreting SMBHs. We propose that these sources are accreting at or beyond the Eddington limit, which reduces the need for efficient production of heavy SMBH seeds at cosmic dawn. Using a theoretical model of super-Eddington accretion, we can produce the observed relative dearth of both X-ray and ultraviolet emission, as well as the high Balmer decrements, without the need for significant dust attenuation. This work indicates that super-Eddington accretion is easily achieved through-out the early Universe, and further study is required to determine what environments are required to trigger this mode of black hole growth. }
    
\maketitle

\section{Main}

The first results from \jwst\ have raised more questions than answers about the prevalence, growth, and impact of accreting supermassive black holes (active galactic nuclei, or AGN) in the first billion years of the Universe. Puzzlingly, a multitude of \jwst\ spectroscopic surveys are finding a significant over-abundance of broad-line AGN (BL AGN) -- two orders of magnitude above predictions at $z>4$ and selected by broadened Balmer emission indicative of $>$ 1000 km $^{-1}$ moving gas  \citep{matthee,maiolino23,harikane23, kocevski23, larson23}. Additionally, a sub-set of these z$\sim5$ \jwst\ BL AGN are characterized by their extreme compactness ( $< 100$ pc) and peculiar ``v-shaped'' SEDs characterized by a puzzling UV excess alongside red \jwst/NIRCam colors ($m_{277} - m_{444} > 1.5$, \cite{barro24}), and are therefore dubbed ``Little Red Dots'' or LRDs \citep{furtak,matthee,greene23,kokorev23, kocevski23,kocevski24,kokorev24,akins24}. While the FWHM of their broadened components is generally narrower than the median FWHM found for bona-fide AGN at lower redshifts (FWHM \halpha\ $\sim$ 2500 km s$^{-1}$, \cite{liu19}; vs 1000 km s$^{-1}$ $<$ FWHM \halpha\ $< 2000$), the relative dearth of significant star-formation signatures points to an accreting SMBH being the likely cause of the rapidly moving gas. 

Several outstanding questions have arisen in attempts to rectify the tension between studies conducted pre- and post-\jwst\ launch. A difficulty in placing \jwst\ spectroscopically-selected BL AGN and photometrically-selected LRDs is the surprising differences between the multi-wavelength observations of these sources and pre-\jwst\ AGN at similar epochs. Despite expectation, many of the BL AGN are not detected in X-rays \citep{maiolino24,ananna24,yue24}, have overall red UV to optical colors with a puzzling blue excess \citep{greene23,kokorev24,akins24} and/or fall short of predictions for their AGN or host-galaxy flux assuming obscuration in the rest-frame near-to-mid infrared and sub-mm \citep{williams24,wang25,akins24,akins25}. This has led to numerous works studying how differences in both AGN properties and/or host galaxy properties could explain why observations of these newly discovered sources lack certain tell-tale signatures of super-massive black hole growth. Some recent studies propose that different dust distributions within these AGN host galaxies could explain the lack of infrared emission \citep{casey24}. Some works suggest that different cloud properties in the broad-line region \citep{maiolino24} or different accretion properties \citep{King_2024,Pacucci_Narayan_2024,yue24} could explain the lack of X-ray detection. Some works have even claimed that unconstrained contributions from star formation in the wavelengths could explain these sources without the need for an AGN at all \citep{williams24,akins24,baggen24}. 

Additionally, over 70\% of z$\sim5$ \jwst\ BL AGN, including spectroscopically confirmed LRDs, have BH masses that are estimated to be at least an order of magnitude above their predicted BH masses via local scaling relations \citep{reines15}, with many over two orders of magnitude \citep{maiolino23,kokorev24}. The implied extremely rapid growth of black holes is challenging to understand. The high ratios of estimated black hole-to-stellar mass of \jwst-discovered AGN alongside the surprisingly massive black holes of very luminous AGN discovered pre-\jwst\ has led to an, at times, controversial emerging picture of early black hole growth: the UV and/or optically luminous black holes we are observing in the early Universe start super-massive (i.e., `massive seeds'), and eventually their host galaxy accretes enough material and/or undergoes a sufficient number of mergers to match local black hole-to-stellar mass relationships \citep{kokorev23,pacucciavi24}. This has led to increased speculation that these over-abundant and over-massive BHs are evidence of efficient direct collapse black hole formation (DCBH) in the early Universe \citep{natarjana24,pacucciavi24,chisholm24}. The masses of these BHs at the time of their observation are large enough where smaller BH seeds would simply not have enough time to accrete sufficient material via moderate to low amounts of accretion (\ledd\ $< 1 $, where \ledd\ is the Eddington ratio). 

This then leads to the question -- what if they were accreting at much higher rates? This scenario, while a common assumption or feature in largely pre-\jwst\ theoretical and simulated studies of early BH growth, was assumed not to play a large role in these newly discovered \jwst\ BL AGN \citep{schneider23}. This is largely because the estimated bolometric luminosities and black hole masses, particularly from the deepest spectroscopic surveys, infer Eddington ratios below unity -- thus implying the mode of accretion would be well-described by sub-Eddington models \citep{maiolino23}. While \jwst\ probes the direct emission from the accretion disk via the rest-frame UV-optical emission, under the conventional models of thin disk accretion, these sources should also be powerful X-ray emitters \citep{signori23, yue24,maiolino24, ananna24}. The connection between the UV/optical accretion disk luminosity and the X-ray emitting region is a powerful probe of the physics governing the entire central engine \citep{lusso17,bisgoni21}. Pre-\jwst\ studies find a remarkably tight correlation between the X-ray-emitting corona and UV/optical accretion disk emission ($\alpha_{\mathrm{OX}}$) for the majority of UV and X-ray selected AGN \citep[e.g.,][]{lusso17}. Yet, there is currently no robust detection of any X-ray emission for any z$>$5 spectroscopically confirmed \jwst\ selected broad-line AGN with anomalous SED shapes. \citep{maiolino23,kocevski23,matthee}. 

Some recent studies have postulated the lack of X-ray emission is either due to 1) these sources not being powered by SMBH accretion at all \citep{pacucciavi24, ananna24,akins24} or 2) the physical broad-line region clouds themselves obscuring the X-ray corona \citep{maiolino24}. While other recent studies have begun to challenge the robustness of the black hole mass estimates used to determine \ledd\ in the first place \citep{bertremes24}. Thus more nuanced approaches are being employed that challenge our assumptions on early BH growth by implicitly testing if we are under-accounting accretion regardless of the reported black hole masses and bolometric luminosities. \cite{lupi24} employ a semi-analytic model approach to show how higher accretion rates and lower black hole masses are preferred, and are even able to recover rest-optical continuum region and broad \halpha properties of a few candidate JWST BL AGN.  This is complemented by simulation based approaches, which find color-selected LRDs are potentially characterized by high Eddington ratios \citep{volonteri25}. Other studies challenge the published \ledd\ values of these new JWST AGN via X-ray stacking. In \cite{yue24}, the significantly weak X-ray detections associated with stacking a diverse sample of the LRD sub-population was explained by associating the relative weakness of X-ray emission predicted in some models of super-Eddington accretion. 

In this study, we aim to concretely determine whether these sources are accreting above the Eddington limit by self-consistently connecting the inconsistencies between predictions and observations across the entirety of the X-ray to NIR regime. We begin by assembling a sample of spectroscopically confirmed serendipitously discovered BL AGN Candidates that have been identified in the literature to have anomalous rest UV-optical SEDs and are covered by deep X-ray observations. We select a sample of 14  BL AGN between 4 $<$ z $<$ 7 that have been previously published from public data and have complete \jwst\ spectral coverage between the rest-UV to optical. Further details on source provenance and selection criteria are given in the Methods.

\begin{figure}[h]
\centering
\includegraphics[width=1\textwidth]{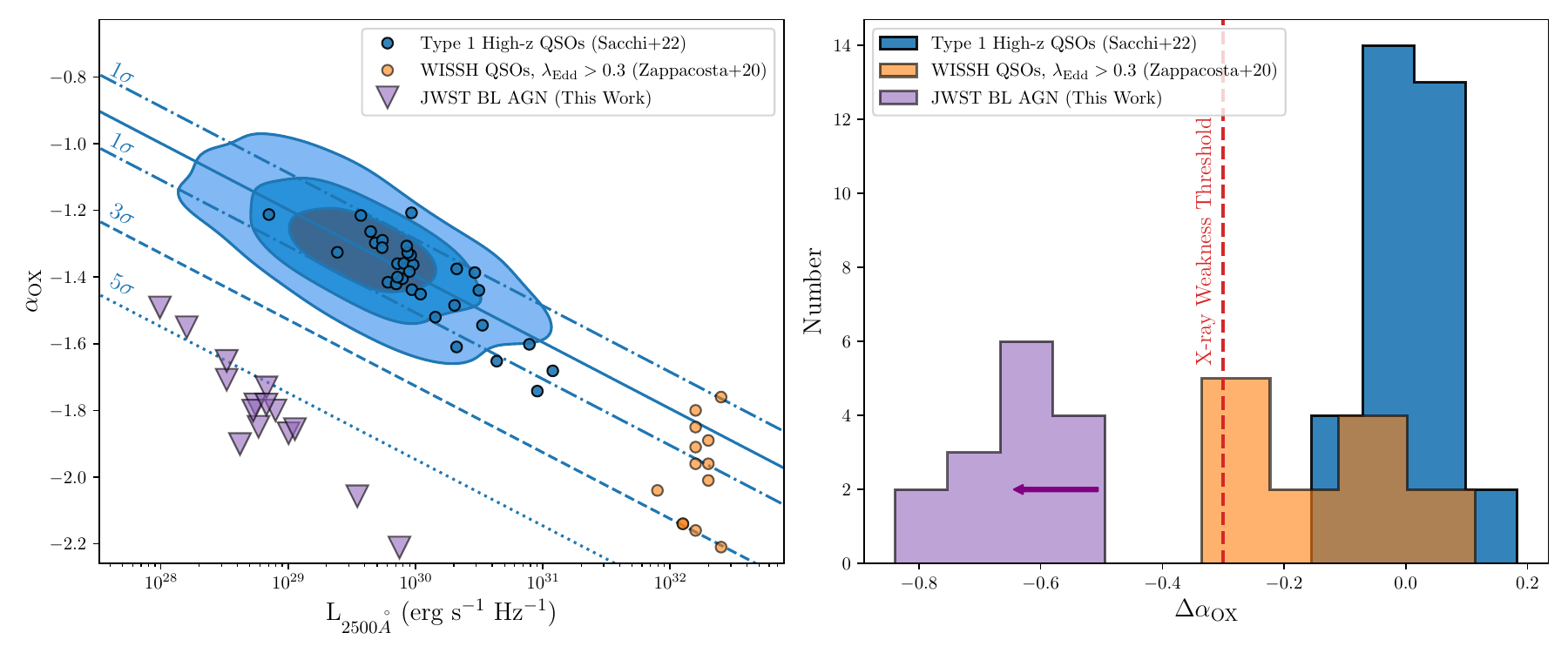}
\caption{Significant X-ray Weakness: (Left Panel) We show the upper limits of \aox\ for the \jwst\ z$\sim5$ BL AGN sample (purple triangles). The blue contours are the spectroscopically-selected BL AGN sample derived from SDSS and confirmed by \cite{lusso17}. The blue solid line shows the \aox\ relationship parameterized by \cite{lusso17}. The dash-dot, dashed, and dotted lines show the 1$\sigma$, 3$\sigma$, and 5$\sigma$ scatter, respectively. The blue points are the high-z Type 1 BL AGN sample (z $> 3$) from \cite{sacchi22}. The orange points are candidate super-Eddington sources from the WISSH QSO sample \cite{zappacosta20}.  (Right Panel) We show the offset from the \cite{lusso17} relationship ($\Delta$\aox) where the colors are consistent with the previous panel. The red dashed line is the canonical X-ray weakness threshold ($\Delta$\aox $< 0.3$, as is shown in \cite{laurenti22}).} \label{fig:aox_figs}
\end{figure}

{\section{A Surprising Lack of X-ray Continuum and UV Line Emission}

We first determine by what margin these sources should have been detected in their respective X-ray surveys assuming the standard sub-Eddington accretion prescription. Assuming these \jwst\ sources are sub-Eddington accretors (as has been done in the discovery papers and subsequent studies on these sources, \cite{maiolino23,matthee}), we employ the widely used empirically derived relationship between the 2 keV and 2500\,\AA\ emission, parametrized as \aox~\citep{lusso17}. Intriguingly, we measure X-ray upper limits that are at least 5$\sigma$ below the predicted \aox\ relation (median $\Delta\alpha_{\mathrm{OX}}=-0.64$). As seen in both panels of Figure \ref{fig:aox_figs}, the upper-limits are well below the scatter of both the local and high-redshift AGN population. Further details are given on the \aox\ parametrization and X-ray/UV/optical data reduction in the Methods section and Table \ref{tab:xrayuls}.  


\subsection{Steeper X-ray Power-Law Slopes are Needed to Explain Non-Detections}

The link between accretion rate ($\dot{\mathrm{M}}$) and the power-law slope of the 0.5-10 keV X-ray spectrum ($\Gamma$) has been well-studied over the past decades \citep{brightman13,trakhenbrot17,liuh21}. The AGN population with the strongest evidence of a significant relationship between $\Gamma$ and \ledd\ are those with high accretion rates \citep{liuh21, Pacucci_Narayan_2024}. The accretion rate is tightly connected to \ledd, where \ledd\ $= L_{\mathrm{Bol}}$/$L_{\mathrm{Edd}}$. \ledd\ is proportional to $\dot{\mathrm{M}}$/M$_{\mathrm{BH}}$ $\propto \dot{\mathrm{M}}$/$\dot{\mathrm{M}}_{\mathrm{Edd}}$, where \mbh\ is the black hole mass. Numerous studies have found evidence that sources with high-\ledd\ tend to have steeper X-ray slopes ($\Gamma > 2$) \citep{liu19}. The most standard explanation that connects higher accretion rate sources with steeper $\Gamma$ is related to the different accretion flows predicted in these sources. The lack of measured X-ray emission is usually ascribed to an intrinsic weakness in the X-ray emission (e.g. photon-trapping where the diffuse timescale for photons to escape from a thick disk surface may be longer than the timescale for photons to be advected into the central black hole) or due to shielding -- both of which have been found to be applicable mechanisms in super-Eddington accretion models. Due to the epochs being probed in this study, the observed-frame 2 keV energies correspond to rest-frame X-ray energies that are $>12$ keV -- thus, any steepening of the power-law from canonical assumptions severely curtails the detectability of these sources in the X-ray. 

While we do not have any X-ray detections to measure $\Gamma$ directly, we can calculate the smallest $\Gamma$ required such that the X-ray emission is not detected within each source's given X-ray observation, assuming the predicted \aox\ value is correct. We note that the empirical \aox\ vs \luv\ relation is contextualized implicitly under models of sub-Eddington accretion, and may be intrinsically different for super-Eddington sources \citep{laurenti22}. We limit this impact by using \luv\ values that are derived red-ward of 2500$\AA$, and thus in a portion of the SED that is predicted to be more similar to sub-Eddington accretion models as opposed to the rest-UV and bluer \citep{leighly07}. We discuss this more in Section \ref{luv}. Using the measured \luv and the canonical \aox\ relation \citep{lusso17}, we use the corresponding monochromatic 2 keV luminosity predicted for these sources to find the minimum $\Gamma$ needed to correspond with the upper-limit X-ray flux of each detection with further details in Section \ref{aoxmethod}. We find a median $\Gamma > 3$, and using these values, we estimate the lower-limit on the corresponding \ledd\ as parametrized in \cite{liuh21} -- yielding a median \ledd $> 2.6$ -- surpassing the Eddington limit. As detailed in Methods \ref{aoxmethod} and shown in Figure \ref{fig:mbh_rates}, we highlight how these new limits reduce the tension between black hole and stellar masses.

\subsection{Intrinsically Missing Rest-UV Ionization Lines}

To concretely test these sources as candidate super-Eddington accretors, we must also inspect other portions of their SED that would be impacted by the intrinsic reduction in X-ray photons -- namely, tracers of high-ionization in the rest-UV. While all these sources are well detected ( $> 5\sigma$) in their respective rest-UV imaging ( $< 2$\micron), it is the spectral rest-UV features that will be most discriminating between competing AGN accretion models. Using the \jwst\ NIRSpec G140M MSA spectra of the sub-set of our sample (7 sources) covered in the JADES survey (see Figures \ref{spectra}, \ref{fig:ha}, Tables \ref{tab:measurements}, \ref{tab:emlines}), we do not find a single source with any detectable UV line emission -- including \civ\ and \ciii, which aside from \lya, comprise some of the brightest lines in the rest-UV at these epochs. In particular, some studies have suggested that the rest-UV SEDs of these sources could be wholly explained through star-formation \citep{kocevski24}. The lack of \ciii, a prominent rest-UV line in low-metallicity SF systems \citep{stark16,mignozzi22,robertsborsani24,tang25} disfavors a SF-only interpretation.} Furthermore, in the case of obscuration as being a factor for the dearth of rest-UV lines, explaining how the much fainter continuum is detected while the brighter emission lines are not is difficult. If the continuum is being driven by AGN emission from the accretion disk, the high-ionization rest-UV lines which are excited in regions beyond the accretion disk cannot be obscured while the rest-UV accretion disk emission isn't. We further explore the impact of obscuration on line detectability Section \ref{sec3}.

Assuming obscuration is the unlikely culprit of the missing UV lines, we turn to explanations that can intrinsically impact the overall properties of the ionizing continuum. In accretion disk models that predict a radiatively inefficient hot corona, the reduced availability of seed photons have been found to have an effect on the nebular regions surrounding the central engine -- namely, the higher ionization potential lines in the broad-line region (BLR) \citep{wu11,jin23wlq}. This suggests that the broad-line emitting gas is incident with an unusually weak  photoionizing continuum. Some studies find that the shape of the ionizing continua in these sources will lead to lower-ionization potential lines, such as the Balmer series, having a less obvious discrepancy from standard accretion models as compared to the higher-ionization potential lines, such as \civ $\lambda 1549 \AA$. Perhaps most relevant to the \jwst\ BL AGN of this study is a similar class of AGN accretors known as weak-line quasars \citep[WLQs;][]{leighly07, diamondstanic09,andika20,jin23wlq}. Pre-\jwst, WLQs were a small fraction of the Type 1 QSO population and are defined by their weak UV high-ionization emission lines; for example, the rest-frame equivalent width of \civ\ $< 10 \AA$ \citep{luo15, ni18}. A substantial fraction of WLQs ($\sim$ 50\%) are X-ray-weak compared to their typical luminous AGN counterparts ($\leq6$\%). In fact, the empirical \aox\ of these sources indicates a significant fraction of the X-ray WLQ population have X-ray fluxes that are mainly undetected or at least a factor of 6 below their expected values \citep{pu2020}. It is also noted that the X-ray weakness was similarly derived with an assumed $\Gamma \sim 2$.

\section{
Missing X-ray and UV Emission under a Super-Eddington Accretion Paradigm} \label{sec3}

Observational evidence of exceeding the Eddington limit is critical in informing the demographics of early Universe black holes. For one, a wide variety of analytical, theoretical, and simulated studies find bursts of super-Eddington accretion are able to reduce the need for efficient massive black hole seed production and/or find the conditions requisite to produce super-Eddington accretion are achievable in the early Universe \citep{lupi16,pezzulli16,pacucci17,regan19,massonneau23,schneider23,lupi24theory,shi24}. The need for different accretion models to describe high-Eddington fraction sources has been known for over thirty years, with many models built off a slim accretion disk or an optically thick, geometrically thicker-than-standard disk with quasi-Keplerian accretion flow \citep{begelman79,abram88,kubotadone19}. These models predict super-Eddington flows that are much less radiatively efficient than standard sub-Eddington discs, as the emitted flux should saturate at the local Eddington limit, with the remaining power lost through radial advection and/or winds. 

With the lack of X-ray detection and the lack of high-ionization UV lines, yet significant broadening of the \halpha\ line component, we construct a model to self-consistently explain these observations under the framework of super-Eddington accretion. For WLQs, some physical interpretations include their broad-line region gas being shielded by a column of gas between the inner-accretion disk and/or X-ray corona and the BLR or an intrinsic X-ray weakness where the corona fails to fully form \citep{wu11,wu24,luo15}. Both of these scenarios are usually attributed to the changes in accretion flow that are expected at higher accretion rates. As previously mentioned, these models relate the inner edge of an optically and geometrically thick ``slim'' disk \citep{abram88,czerny19}. This configuration is expected to shield and/or not intrinsically produce the X-ray and extreme UV radiation incident on the broad-line region gas, thus softening the incident ionizing continuum.


Due to the lack of detection in both UV line emission and X-ray precluding a fit of or sources to models of super-Eddington accretion, we instead compare whether the relative levels of X-ray to NIR emission is consistent with a super-Eddington SED shape. For our approach, we utilize \texttt{agnslim}, the slim disk model as available through \texttt{XSPEC} \citep{kubotadone19}, which adopts a slim disk emissivity -- the surface luminosity is kept at the local Eddington limit within a critical radius. In cases of high-Eddington accretion, the advected flux yields lower fluxes emitted in the UV continuum, and thus an intrinsic reddening of the UV-optical slope is produced. For comparison, we also use the available model in XSPEC, \texttt{agnsed}, to represent a canonical sub-Eddington source with 45$\deg$ inclination angle of the disk \citep{kubotadone18}. For the sub-Eddington case, we construct the SED to reflect how these \jwst\ BL AGN have been previously interpreted in the literature -- namely, AGN with low to modest accretion rates with \mbh\ $\sim 10^{7}$ M$_\odot$ \citep{maiolino23,matthee}. We limit our model comparison to the observed data to the sub-set of sources with G140W observations, and thus set the sub-Eddington SED model parameters with the median \mbh\ and \lbol\ reported in \cite{maiolino23}. 

Since we cannot directly estimate the \lbol\ for these sources under the super-Eddington prescription due to lack of detections, instead we illustrate the relative difference in the predicted X-ray--UV properties between sub- and super- Eddington models. For the super-Eddington case, we normalize the slim disk model to the sub-Eddington model at 4000$\AA$ -- a region of the wavelength space less impacted by the softer ionizing continuum than the rest-UV and still easily comparable to other wavelength bolometric tracers in the literature \citep{shen20}. Additionally, we configure the super-Eddington model parameters with similar values for the electron temperatures, radii for warm and hot comptonization, and outer disk radii to be consistent with the literature for other WLQs with X-ray weakness found in the literature \citep{jin23wlq}. We have set the mass transfer rate relative to \ledd\ = 1, \mbh\ = $10^{6}$ M$_\odot$ and maximal spin. The value of \mbh\ is driven by the upper-limits found in the X-ray non-detection (see Section \ref{aoxmethod}, and the general assumption that for these given sources to be accreting at higher-rates with the same inferred bolometric luminosity derived from the measured \halpha\ flux necessitates a lower \mbh\ than reported by \cite{maiolino23,matthee}. The choice of maximal spin is expected for sources accreting at rates of Eddington or above \citep{inayoshi24}. In Figure \ref{fig:sed}, we show the models as a function of energy. 

\begin{figure}[h]
\centering
\includegraphics[width=0.6\textwidth]{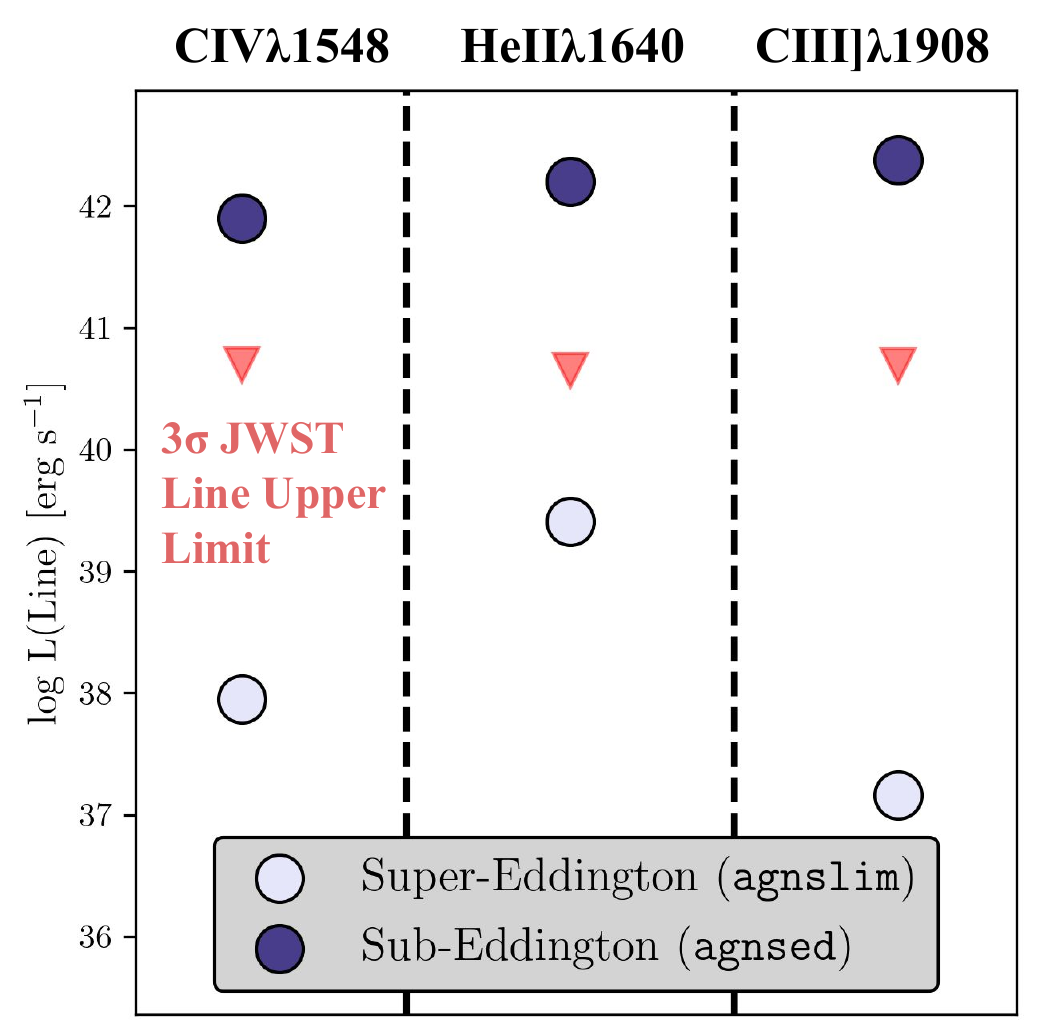}
\caption{Predicted line luminosities for \civ, \heii, and \ciii, respectively. Dark blue is sub-Eddington prescription (agnsed), light purple is slim disk prescription (agnslim). Red triangles are median 3 sigma upper limits from the \jwst/NIRSpec G140M/G235M spectra.}\label{linelum}
\end{figure}

We use these two models as input into \texttt{Cloudy} \citep[v17.02,][]{Ferland17}, see Section \ref{cloudy} in Methods to assess the impact of these different accretion models on the simulated \halpha\ flux and find for both the super-Eddington and sub-Eddington case values consistent with the range of measured \halpha\ fluxes for these sources. The fiducial models for the super-Eddington and sub-Eddington case yield a mean \LHa\ $\sim$ 5$\times10^{41}$ erg s$^{-1}$, \LHa\ $\sim$ 3$\times10^{42}$ erg s$^{-1}$ respectively, which is within the range of the observed broad \LHa\ values reported for the sample \citep{maiolino23}. As shown in Figure \ref{linelum}, we find the \civ, \heii, and \ciii\ line emission is significantly suppressed in the case of slim disk accretion as compared to the canonical sub-Eddington prescription. The line detection limits represent the median 3$\sigma$ line depth at the given line of all the G140M observations with a \cite{calzetti2000} dust law applied. We use the median published values for the G140M observed sub-sample. While there are some dust curves, including the SMC \citep{gordon03}, that can account for the attenuation of UV line emission at these A$_{V}$'s, those same curves would also preclude the UV continuum detections which exist in these samples as well \citep{maiolino23}. Thus, even with accounting for obscuration, the sub-Eddington model predicts we should be able to detect \civ,  \heii, and \ciii\ well above the detection threshold. Even for the maximum published A$_\mathrm{V}$ derived for a source in our sample (A$_\mathrm{V}$ $\sim 1$), the difference between the predicted dust-free \civ\ line luminosity in the sub-Eddington case is over an order of magnitude above the upper-limit of the line detection.

The intrinsic reddening of the X-ray to optical SED of super-Eddington sources as predicted by this model (see Methods, \ref{cloudy}) has significant implications on the interpretation of the rest-UV/optical spectra and photometry as measured by \jwst. Studies that assumed these sources were sub-Eddington accretors via the measured \ledd\ would find the spectro-photometric fitting of these sources with canonical QSO templates attributing the dearth of UV emission due to dust obscuration. Additionally, \jwst\ BL AGN that would also be photometrically selected as ``little red dots'' are found to have Balmer decrement values which, under the assumption of Case B recombination, would indicate significant dust attenuation \citep{kokorev23}. For proof of concept, we measure the Balmer decrement between our fiducial sub-Eddington and super-Eddington accretion models. We find for the sub-Eddington model \halpha/\hbeta\ $\sim$ 3.5, and for the super-Eddington model \halpha/\hbeta\ $\sim$ 8.5. The sub-Eddington model is within the range of normal Balmer decrements of un-obscured sources \citep{baron16}, yet in the super-Eddington case we are well above the expected ratio. It is generally known that the much higher densities of the BLR (as compared to the narrow-line or HII regions) can lead to different Balmer decrements due to line optical depths and collisional effects that are not included in conventional Case B recombination calculations \citep{netzger13,scarlata24}. This is potentially exacerbated in the super-Eddington case, due to the intrinsically weaker ionizing continuum. Thus, without any dust extinction added to the models, a natural consequence of an intrinsically reddened AGN SED would yield higher observed Balmer decrements. 

If indeed these sources are all super-Eddington accretors, we can place an estimate on the duty cycle. Using the 7 sources with rest-UV coverage we compute the number density over the JADES survey area and compare the total number of non-AGN galaxies with similar M$_{\mathrm{UV}} \sim -18.85 [AB]$ at the median redshift of this sample \citep[z$\sim$5,][]{bouwens21}). We find a duty cycle upper-limit $<5\%$. Interestingly, \cite{trinca25} use a semi-analytic model approach to attempt to recover the observed luminosity functions of JWST discovered AGN at z$>$4, and find they required a super-Eddington phase that corresponds to a duty cycle between 0.5 - 4\%, which is consistent with our results. 

\section{Conclusion}

In summary, models of accretion that generate a relative decrease in the EUV ionizing continuum, as is predicted by some models of super-Eddington accretion, decrease the tension between the multi-wavelength properties of high-z \jwst\ BL AGN. These models are able to self-consistently account for 1) the lack of X-ray detection despite sufficiently deep observations, and 2) the lack of UV high-ionization line detection, such as \civ, without the need for high attenuation. Thus, if the true accretion rates of these \jwst\ BL AGN are higher than what is being inferred via the measured \ledd, then it potentially implies that the measured black hole masses of these sources via the FWHM of the broad component of the \halpha\ emission is being over-estimated (since \ledd\ $\propto \mathrm{L}_{\mathrm{Bol}}$/$\mathrm{L}_{\mathrm{Edd}}$ where $\mathrm{L}_{\mathrm{Edd}} \propto \mathrm{M}_{\mathrm{BH}}$). The outstanding question is what fraction of the total LRD and/or serendipitously selected JWST BL AGN population show these hall-mark signs of SE accretion? Thus, applying rest-UV color thresholds and obtaining deep rest-UV spectroscopic follow up is critical to allow for a more expansive study on the total JWST discovered AGN population. We also note that in \cite{maiolino23}, and \cite{matthee}, the bolometric luminosities of the \jwst\ BL AGN are determined via applying a bolometric correction to the luminosity of the broadened \halpha\ line. As is shown in Section \ref{fig:sed}, normalizing the super-Eddington and sub-Eddington accretion models to produce equivalent amounts of \halpha flux will yield significantly different bolometric corrections across the entirety of the multi-wavelength SED. This will significantly impact attempts to accurately predict the amount of rest-frame NIR and MIR flux expected in these sources. For instance, almost every LRD observed between the rest-frame 2-5 $\mu m$ range has significantly less flux than predicted from either energy balance arguments or applying bolometric corrections \citep{wang25,akins24,casey24}. Future work will include applying these accretion models in the context of the measurements of these sources at redder wavelengths.

These results imply that high-Eddington accretors may be more common than currently assumed via observations -- every serendipitous BL AGN found in both the JADES and EIGER-FRESCO survey are consistent with higher Eddington accretion rates as parametrized by our models. This has a significant impact on how we model the growth of the first SMBHs and leaves many avenues of inquiry wide open on connecting the early Universe environment to black hole accretion.  \\

\section{Methods}\label{methods}


\subsection{Parent Sample}
We draw from the published samples that are built off the extensive JWST observations that comprise the \jwst\ Advanced Deep Survey (JADES, \jwst\ GTO 1180, 1181, 1210, 1286, PIs Eisenstein and Luetzgendorf, \cite{bunker23,jades,deugenio24}) and the First Reionization Epoch Spectroscopically Complete Observations (FRESCO) survey (\jwst\ GO 1895, PI Oesch, \cite{oesch23}) which are located within the same observational fields of the deepest Chandra observations ever-performed (GOODS-North: CDFN 2Ms \cite{xue11}, GOODS-South: CDFS 7Ms, \cite{luo17}). The parent AGN samples we select from were presented in \cite{maiolino23} and \cite{matthee}. In addition to deep X-ray observations, we required these sources to be (1) at redshifts $z>4$ to ensure complete photometric coverage from the rest UV to optical wavelengths, (2) were previously characterized as an AGN via a statistically significant ($>$5$\sigma$) broad component ($>1000$ km s$^{-1}$) in \halpha\ and (3) are photometrically characterized as compact with red colors in the \jwst\ NIRCam $>$2 \micron\ imaging bands and significant detections in the $<$2 \micron\ \jwst\ NIRCam imaging bands (see details in \cite{maiolino23, matthee}). These studies comprised some of the first discoveries of these enigmatic AGN candidates with \jwst/NIRSpec and NIRCam grism instruments.  From these surveys, we apply our sample requirements and select 14 rest-frame spectroscopically confirmed BL AGN between 4 $<$ z $<$ 7, seven of which half also have NIRSpec G140W medium resolution spectroscopy to probe their rest-frame UV emission. We stress these sources are most likely not representative of the total AGN demography at these epochs, but rather are some of the most spectroscopically complete and pristine examples of a potentially new class of sources that required \jwst\ for efficient discovery. We note a more complete selection of AGN at this epochs would require access to wavelengths that are less impacted by obscuration, which may be so significant it would preclude detection of the broad-lines in the first place \citep{gilli22, lambrides23}.

\subsection{
\aox and X-ray Upper Limits}\label{xrayul}

In the standard, sub-Eddington accretion paradigm, UV/optical photons from the accretion disk are linked to the X-ray continuum emission that arises via inverse Compton scattering, forming what is conventionally called a ``corona''. The empirical $\alpha_{\mathrm{OX}}$ is the power-law slope that relates the X-ray emitting corona (as probed by the monochromatic rest-frame 2 keV emission) to the accretion disk luminosity (as probed by 2500 \AA\ emission) via \aox\ = $-0.384 \times \mathrm{log}$\,\lxray/\luv. This relationship was found to hold for almost all un-obscured or obscuration-corrected AGN within 0.2 dex scatter -- from local AGN to the highest-z QSOs discovered pre-\jwst\ \citep{lusso17, goudling18,vito2018}. An important note is that at high-enough redshifts, the rest-frame energies being probed by most X-ray telescopes with sufficient sensitivity is much higher than 2 keV (i.e., at z=5, Chandra coverage of 2 keV probes rest-frame 12 keV). Thus the X-ray power-law slope, nominally assumed to be $\Gamma {\sim} 2$ for most AGN at these epochs, is used to infer the rest-frame 2 keV in sources in cases without sufficient counts to measure $\Gamma$ directly.

Using the Chandra Source Catalog V2.1, we downloaded the full-field combined event, arf and rmf stacks of every published Chandra observation that covers each \jwst\ coordinate of the sources used in this sample. These full field combined event files are stacked observation detections event files filtered by the appropriate science energy band. The event files used for the stacked detections event file have been reprocessed through \texttt{acis\_process\_events} to apply the latest instrument calibrations and the standard event status and event grade filters. Additionally, all observations within a given observation stack were aligned and reprojected such that they have a consistent coordinate system. The combined exposure maps are computed by applying the aspect histogram sampled at 0\farcs5 resolution, and are blocked by 1 in SKY coordinates. Robust upper limits were estimated via the exposure-corrected 0.5--2 keV count rates within a 2\arcs\ aperture, and list the 0.5--2 keV flux upper-limits in Table \ref{tab:xrayuls}. The monochromatic 2 kev flux upper limits were determined using the \texttt{CIAO} tools function \texttt{aprates} \citep{ciaotools}, and estimated assuming an X-ray power-law slope of $\Gamma = 2$ and the respective galactic absorbing column density (N$_{\mathrm{Gal}}$ = 8.8$\times 10^{18}$ cm$^{-2}$, 1.1$\times 10^{19}$ cm$^{-2}$ for CDFN/CDFS respectively). Using the redshifts listed in Table \ref{tab1}, we  convert these fluxes to the rest-frame monochromatic 2 keV luminosity and also report the $3\sigma$ upper-limits in Table \ref{tab1}. Finally, we stack all the X-ray observations of our sources. Using the method outlined in \cite{yue24}, built off the Chandra Stacking Tool (CSTACK) as described in \citep{civano16}, we stack in three energy bins (0.5--2 keV, 2--8 keV, and 0.5--8 keV), and do not find a detection in any bin. Specifically we measure a stacked 0.5--2 keV upper-limit of 2.14$\times 10^{-18}$ erg s$^{-1}$ cm$^{-2}$, 90\% confidence.

\subsection{The Connection Between $\Gamma$ and \mbh}\label{aoxmethod}

As stated in the main text, empirical studies have found on average steeper $\Gamma$ in sources accreting above the Eddington limit. To provide a lower-limit on $\Gamma$ we must find how steep does the power-law slope need to be such that predicted monochromatic 2 keV luminosity via \aox\ would be un-detected within the corresponding X-ray surveys depth. Using \aox\ = $-0.384 \times \mathrm{log}$\,\lxray/\luv, and the estimated \luv, we find relatively bright \lxray\ emission expected for these sources under the conventional assumptions of sub-Eddington accretion. To convert these monochromatic luminosities to their respective predicted Chandra 0.5--2 keV fluxes we simply repeat the method used to determine the X-ray upper-limits in reverse. Thus, using the conventional power-law as our spectral input, we solve for the minimum $\Gamma$ needed to re-produce the measured upper-limit Chandra 0.5--2 keV fluxes for each source.

\begin{figure}[h]
\centering
\includegraphics[width=0.75\textwidth]{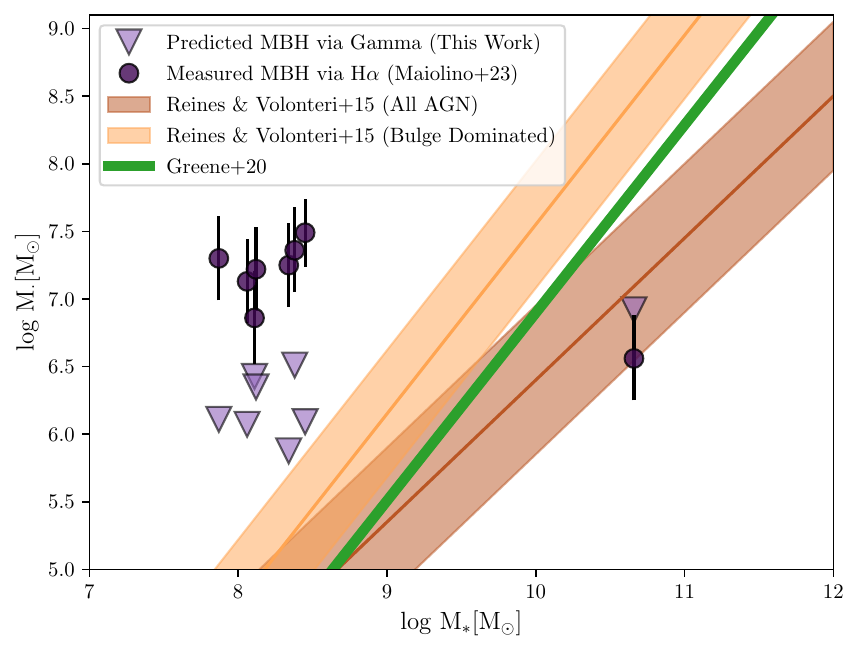}
\caption{Lower than predicted M$_{\mathrm{BH}}$: The dark purple points are the values from \cite{maiolino23} with the \mbh\ errors including both the propagated uncertainty of the measured variable and the scatter of the virial \mbh\ scaling relations. The light orange and brown lines correspond to the relation between black hole and stellar mass for a sample of bulge dominated and the total sample of local AGN respectively, with the width corresponding to the scatter of each relation from \cite{reines15}. The green line is the black hole to stellar mass relationship derived from \cite{greene20}. The light purple triangles are the upper-limits of the black hole mass of our sample that overlaps with \cite{maiolino23}, derived from the upper-limits of the X-ray power slope, $\Gamma$. These black hole mass upper limits are, on average, an order of magnitude below \cite{maiolino23}. }\label{fig:mbh_rates}
\end{figure}

We then use these $\Gamma$ values to estimate the lower-limit on \ledd\ using Eq. (8) in \cite{liuh21}. Finally, we estimate an upper-limit of the black hole mass that is constrained by our X-ray non-detections via \ledd\ $\propto$ \lbol/\mbh\ for the G140M sub-sample with confirmed broad-line features. As shown in \ref{fig:mbh_rates}, using the inferred bolometric luminosities of the sources in our sample \citep{maiolino23,matthee}, our \mbh\ upper-limits are at least an order of magnitude below the literature black hole masses for our sources. In Table \ref{tab1}, we list the associated limits on $\Gamma$ and \mbh, and in Figure \ref{fig:mbh_rates} show the difference between the previously published \mbh\ versus the adjusted \mbh\ when accounting for the estimated \ledd\ derived in this work. We note that the bolometric correction via \halpha\ would be intrinsically different if the sources were accreting above super-Eddington, but in this exercise we simply aimed to highlight that the assumption of uncertainty in robustly measuring \mbh\ at these epochs coupled with anomalously weak X-ray emission can correspond to a higher accretion rates as constrained by the data.

\subsection{An Estimation of the UV Luminosity}\label{luv}

The rest-frame L$_{2500}$ luminosities were estimated via the inferred bolometric luminosity derived from the flux of the broadened \halpha\ component. In the case of the JADES sub-sample we use our re-derived fluxes and compared to published values in \cite{maiolino23}. We found our values to be within the  measurement error reported for each source -- except for MPT ID = 3608. We did not confirm a statistically significant broad component, and thus omit it from the median values used when comparing to the photo-ionization model predictions. In the case of the FRESCO+EIGER sample, we use values published by \cite{matthee}. To be consistent with the literature referenced in our comparisons, we chose to infer the intrinsic L$_{2500}$ emission by probing the reddest available AGN power indicator. Bluer parts of the spectrum which probe rest-frame UV/optical wavelengths are more attenuated by unaccounted for levels of obscuration, and/or more extreme differences in their ionizing continuum under models of super-Eddington accretion. Using the bolometric luminosities inferred from the flux of the broadened \halpha\ component, we apply a bolometric correction via the relation given by \cite{shen20} to estimate the L$_{4000}$ luminosity. Then assuming a canonical f=$\nu^{0.44}$ relation that describes the intrinsic continuum shape of powerful AGN at these wavelengths, infer the L$_{2500}$ luminosity \citep{vandenberk01}.

\subsection{JWST NIRSpec Spectral Extraction and Line Fitting} \label{specmeth}

The sub-sample of our sources (8 objects) with G140M MSA spectral coverage are within the JADES survey. We download the fully-reduced two-dimensional (2D) Spectra of G140M, G235M, and G395M published by the JADES collaborations and detailed in \cite{deugenio24}. These 2D frames were visually inspected for artifacts which were manually masked. We applied an additional sigma clipping with a threshold of 10$\sigma$ to the 2D spectra to remove any additional spurious pixels. Each source's one-dimensional (1D) spectra were obtained via an optimal extraction \citep{Horne86} using a spatial weight profile derived from the trace of the source in the masked 2D spectrum, such that the pixels near the peak of the trace are maximally weighted. To create this extraction profile, we collapsed the 2D signal-to-noise (SNR) spectrum in the spectral direction, taking the median value at each spectral pixel and fitting a Gaussian to the positive trace. 
In the case of source MPT\_ID 1093 there was a second lower-redshift source dispersed into the G140M + G235M spectra due to nearby failed-open shutters during observation. This source prevented a conclusive fit to the profile of the target object in these filters, therefore we used the profile fit obtained from the G395M spectrum to effectively remove this contamination from our final spectrum.  We then combine the spectra from each filter to produce a single spectrum per source covering the full wavelength range of observations. In the regions of wavelength overlap between gratings, we replaced the data from the filter with lower resolution with the higher resolution data. 

\begin{figure}[h!]
\centering
\includegraphics[width=0.86\textwidth]{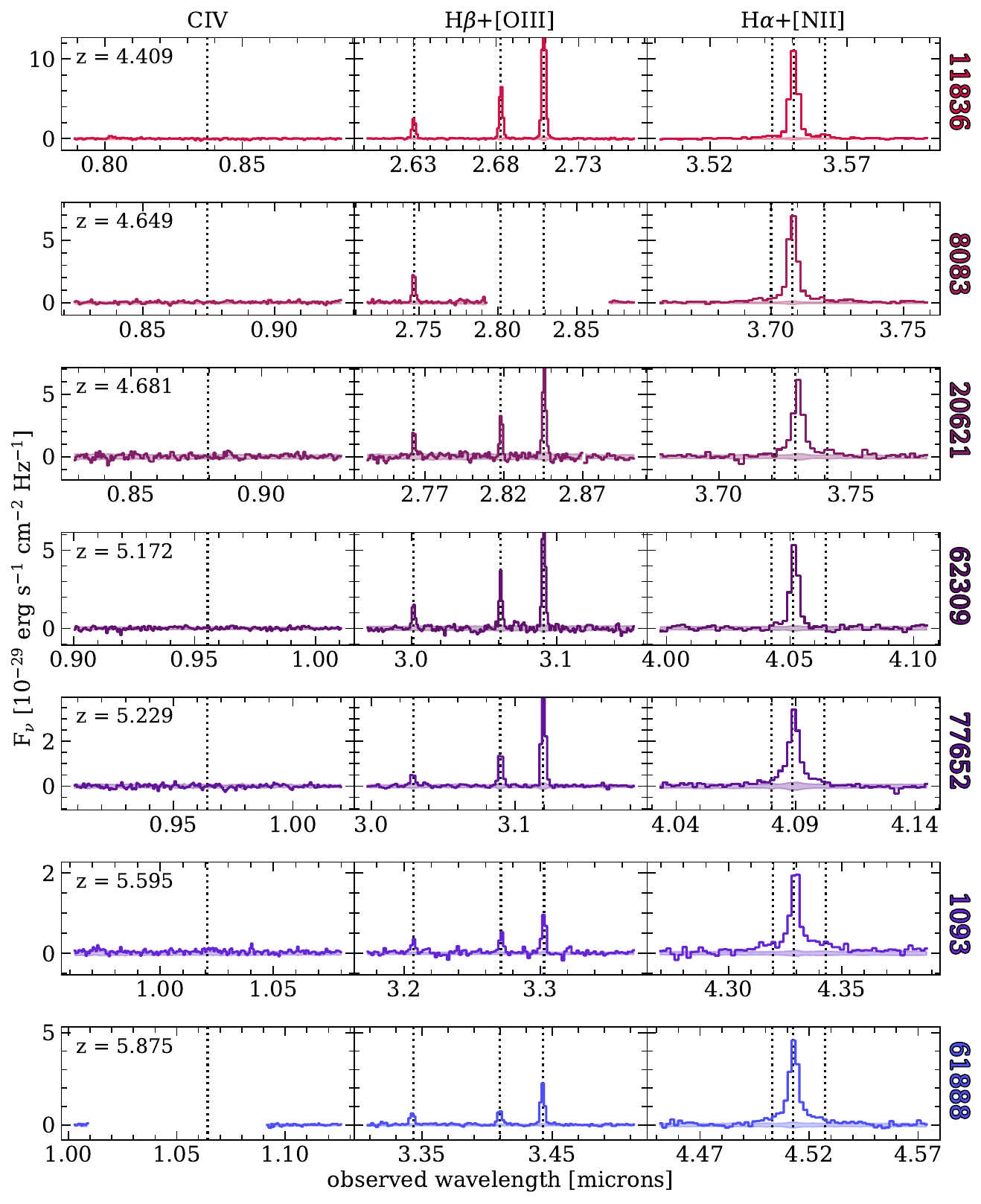}
\caption{\jwst/NIRSpec medium resolution (R$\sim$1000) spectra for the sources with G140M coverage. For each galaxy, the three subpanels show zoomed in regions centered on the expected locations of the \civ, \hbeta\ + \oiii, and \halpha\ + \nii\ emission lines, respectively.  The high-ionization \civ\ emission line is undetected in every source that has wavelength coverage. Details can be found in the Methods section.}\label{spectra}
\end{figure}

As this process removes the neighboring contamination and uses a different extraction method, these 1D spectra differ from those released by the JADES team \citep{deugenio24} and used for the identification of these sources as BL AGN by \cite{maiolino23}. We thus perform our own measurements of the emission lines for consistency in the analysis for this paper. To do so, we ran the 1D spectra through an automated line-fitting routine originally detailed by \cite{larson18} and modified for \jwst/NIRSpec spectra as described by \cite{larson23}. We re-determine the redshift of each source via the \halpha\ line, and identify the \hbeta\ + \oiii\ emission lines, as well as the expected location of \civ. In Figure \ref{spectra}, we show the snippets of our extracted 1D spectra for each source in the sample, marking the locations of these lines of interest. For each of the sources, our redshift measurement is in agreement with the published JADES redshifts \citep{maiolino23}. We note that the JADES public NIRSpec catalog \citep{deugenio24} includes the contamination in the spectrum of MPT\_ID 1093, and emission lines from the interfering galaxy were misidentified as \civ\ emission from the galaxy of interest.

\begin{figure}
\centering
\includegraphics[width=0.9\linewidth]{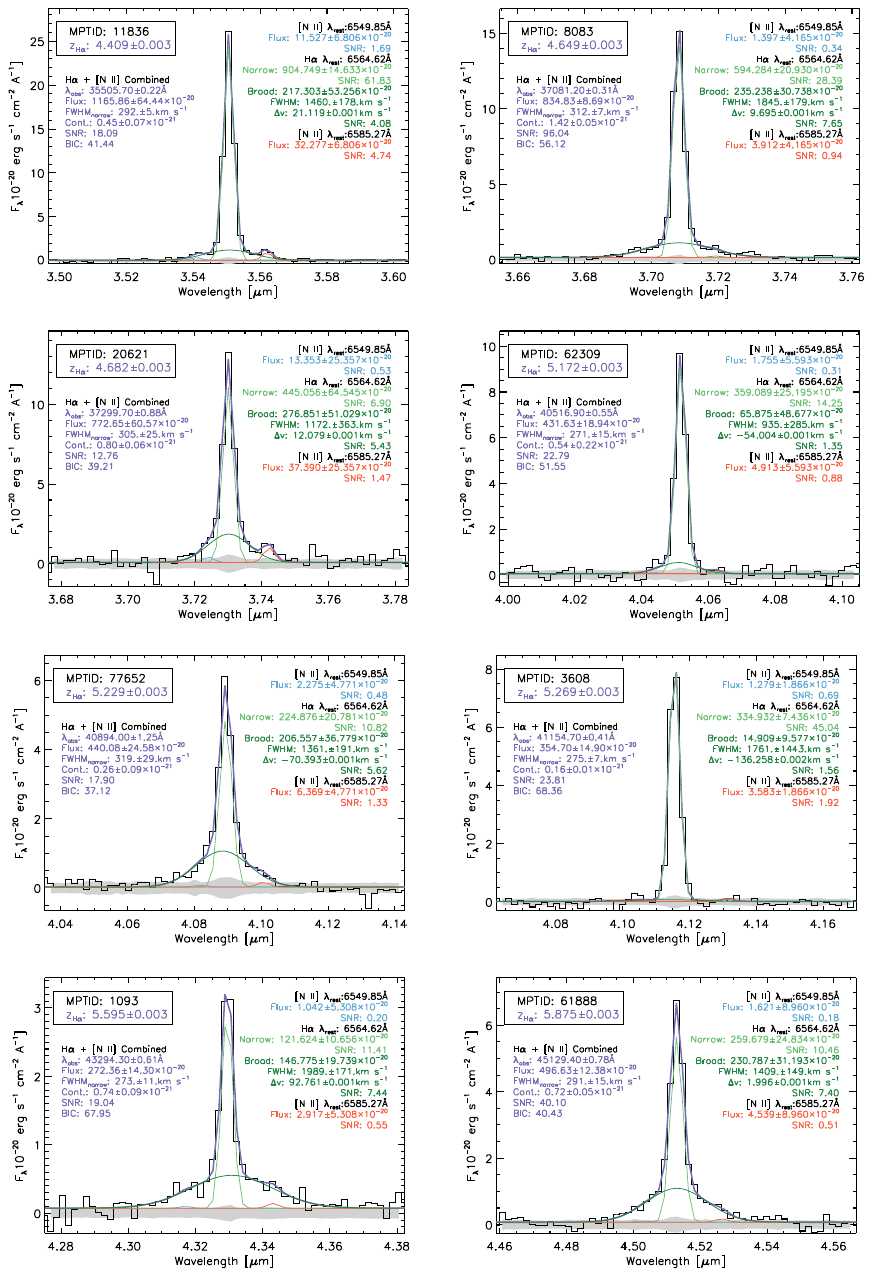}
\caption{Fits to the \halpha\ + \nii\ emission lines for eight sources with NIRSpec/G395M observations in order of increasing redshift. Each is fit with a combination of four Gaussians: narrow components to \halpha\ (light green) and both \nii\ lines (blue and red), plus a broad component to the \halpha\ line (dark green). Best fit parameters for each individual Gaussian are shown on the top right of each plot, and the combined fit (purple) values are shown on the left of each plot. Details are described in the Methods section and values can be found in Table \ref{tab:emlines}.}
\label{fig:ha}
\end{figure}

For each of our target sources we fit the \halpha\ + \nii\ emission line complex as shown in Figure \ref{fig:ha} with a combination of four Gaussians: narrow components  to \halpha\ (light green) and both \nii\ lines (blue and red), plus a broad component to the \halpha\ line (dark green). Best fit parameters for each individual Gaussian are shown on the top right of each plot and reported in Table \ref{tab:emlines}, and the combined fit (purple) values are shown on the left of each plot. The narrow component of the fits are restricted to within 30 km s$^{-1}$  upon the measured FWHM of the \oiii $\lambda 4960+5008$ lines ($\sim$ 300 km s$^{-1}$) and the same FWHM is used for all three narrow lines. In these fits we fix the ratio of \nii$\lambda 6585$/\nii$\lambda 6550$ = 2.8 and restrict the peaks of each to be within 1 pixel ($\sim 18$ \AA) of the redshifted separation from the peak of the \halpha\ emission line. The broad component of the \halpha\ line is not fixed to the same wavelength as the narrow component, allowing for a velocity offset ($\Delta$v), and the FWHM is restricted to $> 3\times$ the narrow FWHM. In most cases, our measured broad FWHM matches that of those reported by \cite{maiolino23} with two notable exceptions. In the case of MPT\_ID 3608, the broad component fit of the \halpha\ line is inconclusive (Figure \ref{fig:ha} b) with a FWHM of 1761 $\pm$ 1443 km s$^{-1}$ and a signal-to-noise ratio (SNR) of 1.56. The broad component fit of the \halpha\ line for MPT\_ID 62309 (Figure \ref{fig:ha}d) is not significant with a SNR of 1.35 and a FWHM of 935 $\pm$ 285 km s$^{-1}$. While MPT\_ID 3608 did make the pre-spectroscopic inspection selection criteria,as there was no conclusive measurement of a broad component it was removed from our BL AGN sample for this paper (and subsequently, not shown in Figure \ref{spectra}). We have indicated so in Table \ref{tab:emlines}.

Once the redshift was established, each spectrum was inspected at the expected location of \civ$\lambda 1548$ for any significant emission feature. With the exception of MPT\_ID 61888, where these wavelengths unfortunately fell in a gap in the data, none of the sources exhibited any significant ($>3\sigma$) feature within 10 pixels ($\sim 60$ \AA) of the location of \civ. In order to get an accurate upper limit measurement we forced a single Gaussian fit at the line location and used the error on that fit as the $1\sigma$ limit. Reported upper limits on the \civ\ emission are 3$\times$ the median measured limits. This same method was used to obtain the $3\sigma$ upper limit values for \heii$\lambda 1640$ and \ciii$\lambda 1907,1909$ for each source. The median of these values for the full sample are shown in Figure \ref{linelum} as red triangles and in Table \ref{tab:emlines}.


\subsection{
Signatures of Sub-Eddington and Super-Eddington Accretion via \texttt{Cloudy} Simulations} \label{cloudy}

\begin{figure}[h]
\centering
\includegraphics[width=0.6\textwidth]{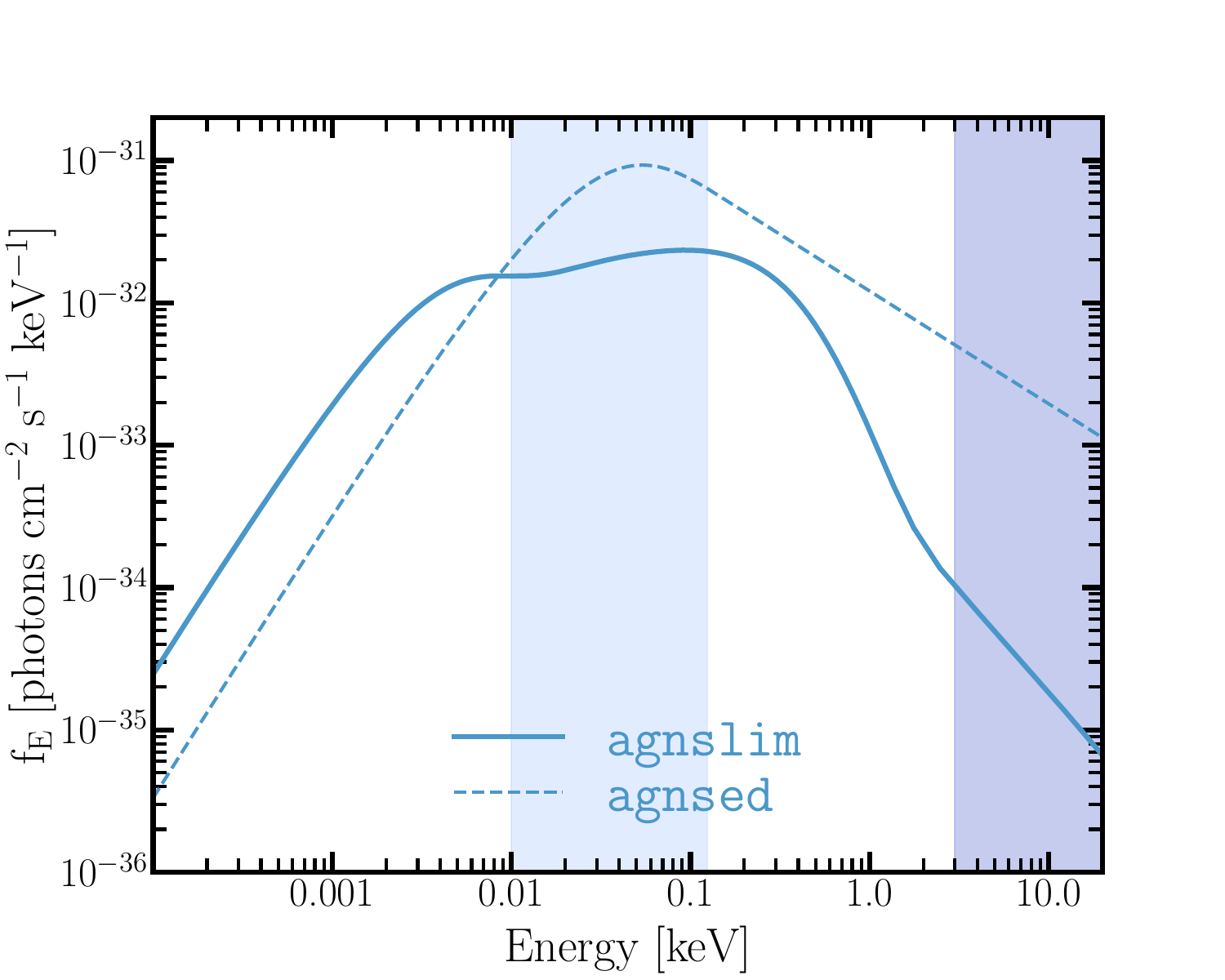}
\caption{Input SED Models for \texttt{Cloudy} normalized by $\mathcal{Q}_{\rm H}$. The solid blue line is the slim disk prescription and the dashed line is a radiatively efficient, sub-Eddington prescription. Light blue box corresponds to the UV-EUV regime, and darker blue box the X-ray regime.} \label{fig:sed}
\end{figure}

For the XSPEC {\tt agnslim} and {\tt agnsed} models, we use radii, temperatures and power-law slopes of all components (i.e., Hot, Warm) as estimated from objects in the literature \citep[e.g.][]{kubotadone18,kubotadone19,jin23wlq} with similar bolometric luminosities (L$_{\mathrm{Bol}} \sim 10^{44.5}$ erg~s$^{-1}$, black hole masses (M$_{\mathrm{BH}} = 10^{6} - 10^{7} \mathrm{M}_{\odot}$), and accretion rates (log $\dot m$ = 0 to $-$0.1) for the super- and sub-Eddington prescriptions respectively. As mentioned in the main text, we conservatively choose maximal spin for {\tt agnslim}. We note assuming a higher black hole spin for this model increases the relative amount of X-ray photons, as compared to a lower black hole spin. Thus, assuming maximal spin is a conservative choice, and any decrease in spin would further decrease the expected amount of X-ray photons. Finally, we note that these selections for the relative shape and normalization ensure \texttt{agnslim} is also consistent with X-ray and UV-continuum upper-limits assuming negligible obscuration. We note, it is important to ensure all parameters in the {\tt agnslim} model are adjusted to be self-consistent with one another. As is seen in Table 1 in \cite{kubotadone19}, the relative difference between the fit values of the super-Eddington candidate RXJ 0439.6-5311 differ significantly from the default values of {\tt agnslim}. In Figure \ref{fig:sed}, we show the Optical--X-ray SED of these physics models for both accretion disk prescriptions, normalized by $\mathcal{Q}_{\rm H}$. We note the intrinsic shape of the slim disk prescription with our model parameters yields a significant difference from the UV to X-ray wavelengths. 

We perform photoionization simulations using the {\tt agnslim} and {\tt agnsed} SEDs as input to \texttt{Cloudy}~v17.02 \citep{Ferland17}. In setting the physical conditions in the cloud, we assume values appropriate to the BLR \citep[e.g.][]{ferland92}. Namely, we assume an open geometry, a fixed inner radius $R \sim 5 \times 10^{16}$~cm, and a density $n_{\rm H} = 10^{10}$ cm$^{-3}$ at the illuminated face of the cloud. Constant gas pressure is assumed until reaching an effective hydrogen column density $N_{\rm H}$ = 10$^{23}$~cm$^{-2}$, at which point the simulations stop. We set the gas-phase abundance to 0.1 $Z_{\odot}$, and use the abundance patterns and scalings from \cite{nicholls17}, where solar is 12 + log(O/H) = 8.76 for the adopted abundance pattern. While most powerful AGN prior to JWST indeed were not extremely metal poor (i.e, generally $>$ 30\% Zsolar) at these epochs, these peculiar sources are consistently below that. For the G140M sub-sample sample, all sources are shifted on the BPT diagram and overlap with metallicity models that range between 10 - 20\% Z$_{\odot}$\citep{maiolino24}. To aid with the use cases of these models, we opted for 10\% to cover a broader range of similar sources at the redshifts of the sample and higher, as the average metallically drops. Finally, we note that dust grains are not included in the cloud.

For the {\tt agnsed} model, we set the ionization parameter to reproduce the median inferred bolometric luminosity of the sample ($\sim$ 7$\times$10$^{44}$ erg~s$^{-1}$). This corresponds to $\log~\mathcal{U}$ = $-2.35$, given the adopted radius and hydrogen density. Note, while we use the ionization parameter to define different ionized regions within the system, we are not basing the run parameters of the models to a fixed ionizing flux. We take radius into account by converting these values into the ionizing photon rate, Q.) We normalize the {\tt agnslim} model to the {\tt agnsed} model at 4000 \AA, such that the inferred bolometric luminosities of these sources via either a bolometric correction to the 4000 \AA~continuum or the BLR H$\alpha$ flux would yield a similar L$_{\mathrm{Bol}}$ from the observers perspective. In reality, the intrinsic bolometric luminosity of the slim disk model would be significantly different -- and thus highlights the need for bolometric corrections that are apt for high-accretion rate systems. This relative normalization yields a slim disk model with an ionizing photon rate ($\mathcal{Q}_{\rm H}$) that is $\sim$ 40$\times$ lower than the ionizing photon rate for the {\tt agnsed} model. As a result, the normalization for the {\tt agnslim} model in \texttt{Cloudy} corresponds to $\log~\mathcal{U}$ = $-$4 which we note is smaller than the canonical BLR value of sub-Eddington sources.

\bmhead{Data Availability}
The fully reduced 2D spectroscopic JWST data used in this work are from the JADES Public Data Release and may be obtained from the MAST archive at \cite{jadesdoi}. The corresponding ID numbers for JADES GOODS-North, identified in `Catalog ID' of Table \ref{tab:xrayuls} is a combination of the survey name `goods-n-mediumhst-0000' with the ID listed in `NIRSpec MPT\_ID' in Table \ref{tab:measurements}. The one JADES GOODS-S spectrum used is with the survey named `goods-s-deephst'. The combination of disperser and filter used is both  `f070lp-g140m' and "f290lp-g395m" for the G140M and G395M spectra respectively with version number `v1.0'. 

\bmhead{Acknowledgments} 

We thank the referees for the helpful feedback in improving the manuscript. We thank A. Sacchi for discussions surrounding the different approaches in connecting X-ray upper limits to competing accretion disk models. E.L., K.G., and T.A.H. are supported by appointment to the NASA Postdoctoral Program (NPP) at NASA Goddard Space Flight Center, administered by Oak Ridge Associated Universities under contract with NASA.  R.L.L. appreciates support from a Giacconi Fellowship at the Space Telescope Science Institute, which is operated by the Association of Universities for Research in Astronomy, Inc., under NASA contracts NAS 5-26555 and NAS5-03127.
This work is based upon observation made with the NASA/ESA/CSA James Webb Space Telescope. The data were obtained from the Mikulski Archive for Space Telescopes at the Space Telescope Science Institute, which is operated by the Association of Universities for Research in Astronomy, Inc., under NASA contract NAS 5-03127 for JWST.
\bmhead{Statement of Competing Interests}
The authors declare no competing interests.

\bmhead{Author Contributions}

E.L. led the full analysis and scientific interpretation for this work.  R.L.L. and K.G. contributed equally to this work, supporting the first author in various stages of the analysis and data reduction. A.P., M.C., A.S.L., T.A.H., C.N., J.M., H.B.A., D.A.B., J.C., F.C., A.P.C., R.E., A.L.F., R.G., S.G., M.H., J.S.K., D.D.K., V.K., F.P., C.T.R., M.S., and K.E.W provided feedback during analysis, scientific interpretation, and/or writing of this manuscript.



\clearpage

\begin{table}

    \centering
    \caption{Chandra 0.5--2keV upper limits at the  90\% confidence level.}    
    \begin{tabular}{lc}
    \toprule
    \multirow{2}{*}{Catalog ID} & $f_{0.5-2keV}$ \\
    & {\footnotesize [erg/s/cm$^2$]} \\ 
    ~\vspace{-3mm}\\ 
    \hline
    ~\vspace{-2mm}\\ 
       JADESGN189220596226368  & $<$\,7.30$\times10^{-18}$   \\
       JADESGS53132842780186  & $<$\,7.2$\times10^{-18}$  \\
       JADESGN189122526229285  & $<$\,1.15$\times10^{-17}$  \\
       JADESGN189248986221835  & $<$\,1.68$\times10^{-17}$  \\
       JADESGN1892932362199 & $<$\,1.85$\times10^{-17}$   \\
       JADESGN18911794223552  & $<$\,3.06$\times10^{-17}$  \\
       JADESGN189179746222463  & $<$\,6.74$\times10^{-18}$  \\
       JADESGN189168026221701  & $<$\,9.81$\times10^{-18}$   \\
       GOODS-N-4014   & $<$\,1.65$\times10^{-17}$ \\
       GOODS-N-9771  & $<$\,2.14$\times10^{-17}$ \\
       GOODS-N-12839  & $<$\,2.11$\times10^{-17}$ \\
       GOODS-N-13733  & $<$\,2.50$\times10^{-17}$ \\
       GOODS-N-14409  & $<$\,1.14$\times10^{-17}$ \\
       GOODS-N-15498  & $<$\,1.30$\times10^{-17}$ \\
       GOODS-S-13971  & $<$\,4.20$\times10^{-18}$  \\
       ~\vspace{-4mm}\\ 

    \botrule   
    \end{tabular}

    \label{tab:xrayuls}
    ~\vspace{-10mm}\\ 
   
\end{table}

\begin{table}[h]
\caption{The measured and published values for the sources in our sample: The \jwst/NIRSpec observations of these sources were taken from JADES \citep{deugenio24} and \jwst/NIRCam grism observations from EIGER \citep{matthee}. }\label{tab1}%
\begin{tabular}{@{}p{3.5cm}p{1.3cm}p{1.3cm}p{1cm}p{1.4cm} p{.95cm}p{.8cm}p{1.5cm}p{.9cm}p{1cm}@{}}
\toprule
  & RA & Dec & Redshift\footnotemark[1] & L$_{2 \mathrm{kev}}$ & $\Delta$\aox\ & $\Gamma $&  M$_{\mathrm{BH}}$\footnotemark[1] & M$_{\mathrm{BH}}$&  NIRSpec \\
JADES Catalog ID  & J2000 & J2000 & $z$ &  erg/s &  & & log(M$_\odot$) & (This Work) &  MPT\_ID  \\
\midrule
JADESGN189220596226368 & 189.22059 & 62.26368& 4.40935 &$<$ 2.2e42 & $<$ -0.61 & $>2.8$ &7.13$^{+0.31}_{+0.31}$ & $<$ 6.07 & 11836\\
JADESGS53132842780186 & 53.13284  & -27.80186& 4.6482  &$<$ 1.2e42 & $<$ -0.78 & $>3.5$  &7.49$^{+0.31}_{+0.31}$& $<$ 6.09 & 8083 \\
JADESGN189122526229285 & 189.12252 & 62.29285& 4.68123 &$<$ 2.1e42& $<$ -0.66 & $>3.2$&7.30$^{+0.31}_{+0.31}$  & $<$ 6.1 & 20621\\
JADESGN189248986221835 & 189.24898 & 62.21835& 5.17241 &$<$ 3.0e42  & $<$ -0.51 & $>2.1$ &6.56$^{+0.32}_{+0.31}$  & $<$ 6.92 &62309\\
JADESGN1892932362199   & 189.29323 & 62.199& 5.22943 &$<$ 3.4e42 & $<$ -0.55 & $>2.5$  &6.86$^{+0.35}_{+0.34}$  & $<$ 6.42 &77652\\
JADESGN18911794223552  & 189.11794 & 62.23552& 5.26894 &$<$ 2.7e42 & $<$ -0.50 & $>3.1$&6.82$^{+0.38}_{+0.33}$  & $<$ 6.03 & 3608\footnotemark[2] \\
JADESGN189179746222463 & 189.17974 & 62.22463& 5.5951  &$<$ 4.6e42 &$<$ -0.57 & $>2.6$ &7.36$^{+0.32}_{+0.31}$  & $<$ 6.51 &1093\\
JADESGN189168026221701 & 189.16802 & 62.21701& 5.87461 &$<$ 3.5e42 & $<$ -0.62 & $>2.2$ &7.22$^{+0.31}_{+0.31}$  & $<$ 6.34 &61888\\
\toprule
 &   RA & Dec & Redshift\footnotemark[3]  & L$_{2 \mathrm{kev}}$ & $\Delta$\aox\ & $\Gamma $&  M$_{\mathrm{BH}}$\footnotemark[1] & M$_{\mathrm{BH}}$ \\
 FRESCO Catalog ID & J2000 & J2000 & $z$&  erg/s &  & & log(M$_\odot$) & (This Work)  \\
\midrule
GOODS-N-4014  & 189.30013 & 62.21204 & 5.228& $<$ 2.9e42 & $<$ -0.67 & $>3.5$ & 7.58 $\pm 0.08$ & $<$ 6.32\\
GOODS-N-9771  & 189.28100 & 62.24730 & 5.538& $<$ 2.8e42 &$<$ -0.84& $>7.3$ & 8.55 $\pm 0.03$  & $<$6.85\\
GOODS-N-12839 & 189.34429 & 62.26336 & 5.241& $<$ 3.2e42& $<$ -0.75 & $>5.6$ &8.01 $\pm 0.06$ & $<$ 6.64\\
GOODS-N-13733 & 189.05708 & 62.26894 & 5.236& $<$ 2.6e42& $<$ -0.64 & $>2.6$&7.49 $\pm 0.10$  & $<$6.20\\
GOODS-N-14409 & 189.07208 & 62.27343 & 5.139& $<$ 3.4e42& $<$ -0.62 &$>2.6$ &7.21 $\pm 0.14$  & $<$ 6.34\\
GOODS-N-15498 & 189.28554 & 62.28078 & 5.086& $<$ 3.4e42&$<$ -0.65 & $>3.3$ &7.71 $\pm 0.11$& $<$6.40\\
GOODS-S-13971 & 53.138583 & -27.79025 & 5.481& $<$ 1.9e42& $<$ -0.70 & $>3.5$ &7.49 $\pm 0.25$ & $<$ 6.09\\

\botrule
\label{tab:measurements}
\end{tabular}
\footnotetext{ $^{1}$Published values from \cite{maiolino23}.\\ $^{2}$ Despite matching the initial pre-spectroscopic inspection selection criteria, this object was removed from our sample for this paper due to an inconclusive/non-detection of a broad line feature in the \halpha\ emission line.\\ $^{3}$ Published values from \cite{matthee}. }
\end{table}

\begin{table}[h] \label{tab:emlines}
\caption{The emission line measurements for the sources in the sample with NIRSpec M-grating (R$\sim1000$) observations. Line fluxes and $3\sigma$ upper limits are presented in units of 10$^{-19}$ erg s$^{-1}$ cm$^{-2}$ \AA$^{-1}$ and FWHMs and $\Delta $v in units of km s$^{-1}$. Description of fitting methods can be found in the Methods section and plots of each fit are shown in Figures \ref{fig:ha} and \ref{fig:ha}. Missing values are due to gaps in the spectral coverage for that source. }\label{tab:emlines}%

\setlength{\tabcolsep}{3pt} 

\begin{tabular}{@{}lcccccccccc@{}}
\toprule

NIRSpec & Narrow Line & \halpha$\lambda 6563$  & \nii$\lambda 6550$  & \nii$\lambda 6585$ & Broad \halpha & Broad \halpha & Broad & \civ  & \heii & \nev \\
MPT\_ID & FWHM  & Line Flux & Line Flux & Line Flux & Line Flux &  FWHM & $\Delta$v & $3\sigma$  & $3\sigma$  &  $3\sigma$ \\
\midrule
11836 & 291.52 $\pm$ 5.24 & 90.47 $\pm$ 1.46 & 1.15 $\pm$ 0.68 & 3.23 $\pm$ 0.68 & 21.76 $\pm$ 5.33 & 1460.00 $\pm$ 178.13 & 21.12 & $<$ 0.9 & $<$ 8.0 & - \\
8083 & 312.47 $\pm$ 6.55 & 59.43 $\pm$ 2.09 & 0.14 $\pm$ 0.42 & 0.39 $\pm$ 0.42 & 23.50 $\pm$ 3.07 & 1844.67 $\pm$ 178.80 & 9.70 & $<$ 5.1 & $<$ 6.2 & $<$ 4.4 \\
20621 & 305.36 $\pm$ 24.68 & 44.51 $\pm$ 6.45 & 1.34 $\pm$ 2.54 & 3.74 $\pm$ 2.54 & 27.69 $\pm$ 5.10 & 1172.29 $\pm$ 363.01 & 12.08 & $<$ 16.7 & $<$ 11.9 & $<$ 0.1 \\
62309 & 270.84 $\pm$ 14.81 & 35.91 $\pm$ 2.52 & 0.18 $\pm$ 0.56 & 0.49 $\pm$ 0.56 & 6.59 $\pm$ 4.87 & 935.35 $\pm$ 285.46 & 54.00 & $<$ 21.6 & $<$ 16.1 & $<$ 1.5 \\
77652 & 319.33 $\pm$ 29.36 & 22.49 $\pm$ 2.08 & 0.23 $\pm$ 0.48 & 0.64 $\pm$ 0.48 & 20.66 $\pm$ 3.68 & 1361.00 $\pm$ 191.36 & -70.39 & $<$ 0.6 & $<$ 0.6 & $<$ 4.2 \\
3608\footnotemark[1] & 275.25 $\pm$ 6.88 & 33.49 $\pm$ 0.74 & 0.13 $\pm$ 0.19 & 0.36 $\pm$ 0.19 & 1.49 $\pm$ 0.96 & 1760.62 $\pm$ 1443.46 & -136.26 & $<$ 13.1 & $<$ 9.2 & $<$ 3.1 \\
1093 & 273.38 $\pm$ 11.02 & 12.16 $\pm$ 1.07 & 0.10 $\pm$ 0.53 & 0.29 $\pm$ 0.53 & 14.68 $\pm$ 1.97 & 1988.64 $\pm$ 170.75 & 92.76 & $<$ 7.6 & $<$ 3.7 & $<$ 1.5 \\
61888 & 290.69 $\pm$ 14.79 & 25.98 $\pm$ 2.48 & 0.16 $\pm$ 0.90 & 0.45 $\pm$ 0.90 & 23.08 $\pm$ 3.12 & 1409.37 $\pm$ 149.03 & 2.00 & - & $<$ 5.7 & $<$ 2.8 \\
    
\botrule
\end{tabular}
\footnotetext{ $^{1}$This object was removed from our sample for this paper due to an inconclusive/non-detection of a broad line feature in the Ha emission line.}
\end{table}

\end{document}